\documentclass[12pt]{iopart}

\usepackage{iopams}

\usepackage{graphicx,color}
\newtheorem{asp}{Assumption}

\begin{document}

\title{Exact Analysis of the Response of Quantum Systems to Two Photons Using a QSDE Approach}

\author{Yu Pan}

\address{School of Engineering and Information Technology, University of New South Wales, Canberra, ACT, 2600, Australia}

\author{Daoyi Dong}

\address{School of Engineering and Information Technology, University of New South Wales, Canberra, ACT, 2600, Australia}

\author{Guofeng Zhang}

\address{Department of Applied Mathematics, The Hong Kong Polytechnic University, Hong Kong, China}
\ead{Guofeng.Zhang@polyu.edu.hk}

\vspace{10pt}
\begin{indented}
\item[]September 2015
\end{indented}

\begin{abstract}
We introduce the quantum stochastic differential equation (QSDE) approach to exactly analyze the response of quantum systems to a continuous-mode two-photon input. The QSDE description of the two-photon process allows us to integrate the input-output analysis with the quantum network theory, and so the analytical computability of the output state of a general quantum system can be addressed within this framework. We show that the time-domain two-photon output states can be exactly calculated for a large class of quantum systems including passive linear networks, optomechanical oscillators and two-level emitter in waveguide systems. In particular, we propose to utilize the results for the exact simulation of the stimulated emission as well as the study of the scattering of two-mode photon wave packets.

\end{abstract}

%
%
%
%
%

\section{Introduction}
The study of the dynamics of photon-photon interaction is fundamental in physics. Particularly, the controllable photon-photon interaction may play a vital role in the realization of all-optical circuits and quantum information processing \cite{Chang07,Chen13,Tiecke14}. For example, the transmission of single-photon signal might be controlled by a gated photon, leading to a novel design of photonic transistor which operates with minimum energy usage \cite{Chang07}. Since photons rarely interact in free space, quantum systems are often employed to mediate the interaction. There exist numerous proposals for the mediation of light-light interaction using quantum systems such as artificial atoms in waveguides and molecules \cite{Shen07,Neumeier13,Hwang09}.

In the language of system theory, the mediated two-photon interaction can be understood as the response of a quantum control system to a two-photon input. The output state carries the full information of the response of the system, which can be further used for photon statistics and correlation analysis. The exact calculation of the two-photon response of a two-level system has been studied using input-output formalism \cite{Fan10}, Bethe-ansatz method \cite{Shen07} and Lehmann-Symanzik-Zimmermann reduction \cite{Shi11}. The generalized treatment of photon-photon interaction has also been studied in \cite{Xu13,Xu15}, focusing on the analytical property of the scattering matrix. A diagrammatic approach is studied in \cite{PhysRevLett.113.183601} which took into account the effect of the relaxation of two distant qubits in scattering. In general, these methods model the interaction between the photons and the system as an inelastic scattering process, which could facilitate the stationary state analysis in either frequency or time domain \cite{Rep12,Liao13}. The multi-photon response of linear (harmonic oscillators) and finite-level systems has also been investigated in \cite{Bara12,Guofeng14} using QSDE equations. These works have shown that some important quantities such as output photon flux and covariance function can be conveniently calculated using the QSDE equations. Generally speaking, it is straightforward to study the time-domain dynamics by solving the Schr\"{o}dinger equation which governs the interaction between the photon wave packets and the control system \cite{Valente12,Nysteen14}. For example, numerical results for the scattering of photon wave packets by a two-level emitter in one-dimensional waveguide have been obtained using this wave function approach \cite{Nysteen14}. Despite the progress, however, the exact calculation of the time-domain output state is still challenging due to the complexity of two-photon dynamics.

In this paper we propose a QSDE approach for the general modelling of the two-photon process. QSDE \cite{Parth1992} is the generalization of the equation of motion in Heisenberg picture, which can characterize the evolution of the operators for a general quantum system or a coherent quantum network. Specifically, QSDE can be derived using the parameters of the overall system, while the Hamiltonian and field coupling operators of the overall system are calculated according to the interconnection between the subsystems \cite{gardiner04,Gough09,James10}. Therefore, the QSDE approach is capable of dealing with an integrated quantum system which may involve arbitrary number of subsystems \cite{Hendra09,Zhang12,Tezak5270}. Moreover, it is convenient to utilize the QSDE approach to calculate the analytical form of the time-domain output state. As we will show in Section \ref{secpl}, the time-domain output state can be analytically calculated for a general passive linear network. The exact two-photon response of an optomechanical system and a two-level emitter can be exactly calculated as well.

The QSDE formalism for dealing with two-photon response of a general quantum system is developed in Section \ref{secio}. The analytical computability of the output state is discussed in Section \ref{AC}. In Sections \ref{secpl}, \ref{secop} and \ref{sectl}, we show some applications of the QSDE approach. Particularly, we consider the simulation of the stimulated emission and the scattering of two photons in a waveguide, based on the exact calculation of the output state. Conclusion is presented in Section \ref{seccon}.

\section{The input-output formalism for uncorrelated and general two-photon input states}\label{secio}
\begin{figure}
\includegraphics[scale=0.6]{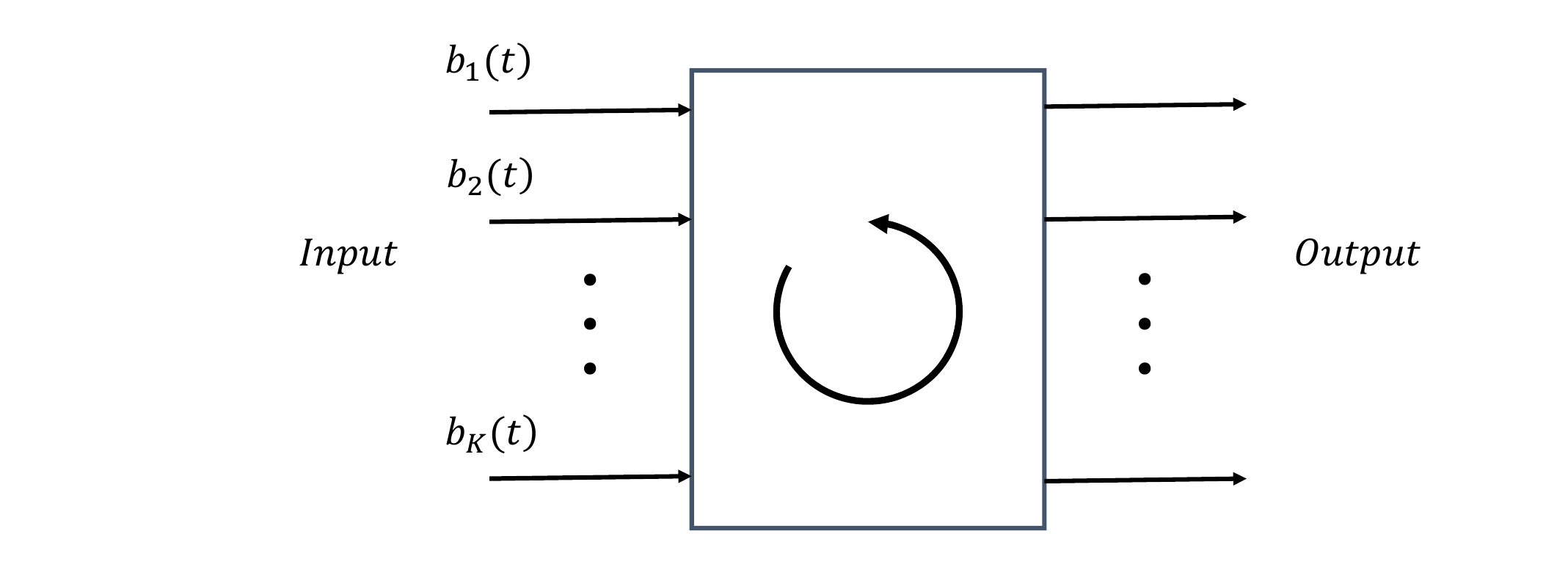}
\caption{A quantum system may have multiple input channels. The Hamiltonian and the coupling operators of the total system are determined by its internal structure, which may involve interconnection of subsystems.}
\label{figmul}
\end{figure}

In this section, we propose a formalism to model the interaction between a continuous-mode two-photon input and a general quantum system.

First, we introduce the QSDE description of the dynamics of open quantum systems. The dynamics of an open quantum system interacting with the input fields is generated by a unitary evolution characterized by a unitary operator $U(t,t_0)$, where $t_0$ is the initial time of the interaction. The dynamical equation of $U(t,t_0)$ is given by
\begin{equation}\label{lem1p}
dU(t,t_0)=\{b^\dag(t)L-L^\dag b(t)-(\frac{1}{2}L^\dag L+\mbox{i}H_0)\}U(t,t_0)dt,\ t\geq t_0,
\end{equation}
with $U(t+dt,t_0)=U(t,t_0)+dU(t,t_0)$ and $U(t_0,t_0)=I\otimes I$. The system is coupled to the external fields through $K$ channels (Figure~\ref{figmul}), with the coupling operator defined by $L=[L_1\cdot\cdot\cdot L_K]^T$ which is a column vector of operators. $H_0$ is the system Hamiltonian. $b(t)=[b_1(t)\ \cdots \ b_K(t)]^T$ is a column vector of bosonic field annihilation operators, and $L_i,b_i(t),i=1,2,\cdot\cdot\cdot,K$ are defined on the $i$-th channel. The singular field operators satisfy the commutation relation $[b_i(t),b_j^\dagger(s)]=\delta(t-s),\ i=j$ and $[b_i(t),b_j^\dagger(s)]=0,\ i\neq j$. Formally, $B_i(t)=\int_0^tb_i(s)ds$ is the quantum Wiener process and $dB_i(t)=B_i(t+dt)-B_i(t)$ is the operator-valued Ito increment. Eq.~(\ref{lem1p}) is obtained by modelling the environment Hamiltonian as $\int_{-\infty}^\infty\omega b^\dag(\omega)b(\omega)d\omega$, and the interaction Hamiltonian as $\int_{-\infty}^\infty(L^\dag b(\omega)+b^\dag(\omega)L)d\omega$ with $L$ being independent of $\omega$. Here rotating wave and Markov approximations have been invoked to obtain the current solvable form of interaction Hamiltonian. It is worth mentioning that the rotating wave and Markov approximations are generally valid for quantum photonic systems \cite{gardiner04,Fan10}, where interaction strength is relatively low, the incident photons are near resonance with the system transition frequency and the environment has no memory effects. The systems considered in this paper can be operated within this regime.

Please also note that the choice of stochastic calculus would not affect the calculations of physical quantities \cite{Gough99ca}.

Denote $|0\rangle$ as the vacuum field state and $|0_s\rangle$ as the ground state of the system. Throughout this paper we concern with systems possessing the simple passivity property
\begin{equation}\label{pas1}
U(t,t_0)|00_s\rangle=|00_s\rangle,
\end{equation}
by which we can prove
\begin{equation}\label{pas2}
U^\dag(t,t_0)|00_s\rangle=U^\dag(t,t_0)U(t,t_0)|00_s\rangle=|00_s\rangle.
\end{equation}
The condition Eq.~(\ref{pas1}) can be easily established as long as $H_0$ and the couplings add no energy to the overall system. For later use, we make the following assumption throughout this paper:
\begin{asp}\label{assumption1}
The number of quanta in the overall system is a conserved quantity as time evolves.
\end{asp}
Under this assumption, the system will remain at the ground state when initially the system is at the ground state and the field is vacuum. The Heisenberg-picture evolution of a system operator $X$ is defined by $X(t)=U^\dagger(t,t_0)(I\otimes X)U(t,t_0)$, with $I$ being the identity operator on the field. Based upon Eq.~(\ref{lem1p}), the dynamical equations of $X(t)$ are derived as QSDEs \cite{hudson84,PhysRevA.31.3761,Parth1992,Gough12}:
\begin{eqnarray}
\dot{X}(t)&=&\mathcal L^\dagger(X(t))+b^\dagger(t)[X(t),L(t)]+[L^\dagger(t),X(t)]b(t),\label{nopr1}\\
b_{out}(t)&=&L(t)+b(t),\label{nopr2}
\end{eqnarray}
where the generator $\mathcal L^\dagger(X(t))$ is given by
\begin{eqnarray}\label{pn1}
&&\mathcal L^\dagger(X(t)):=-\mbox{i}[X(t),H_0(t)]\nonumber\\
&+&\sum_{k=1}^K(L_k^\dagger(t) X(t)L_k(t)-\frac{1}{2}L_k(t)^\dagger L_k(t)X(t)-\frac{1}{2}X(t)L_k^\dagger(t) L_k(t)).
\end{eqnarray}
It is clear from Eq.~(\ref{nopr1}) that the evolution of a system operator is driven by the input field. Moreover, the output field operator $b_{out}(t)$ is related to the input field operator $b(t)$ by the following relation \cite{Gough12}
\begin{equation}\label{inout}
b_{out}(t)=U^\dagger(t,t_0)b(t)U(t,t_0)=U^\dagger(\tau,t_0)b(t)U(\tau,t_0),\ \tau\geq t.
\end{equation}
That is, the unitary evolution transforms $b(t)$ into $b_{out}(t)$ in an infinitesimal time interval, which is a consequence of the Markov approximation.

\begin{figure}
\includegraphics[scale=0.6]{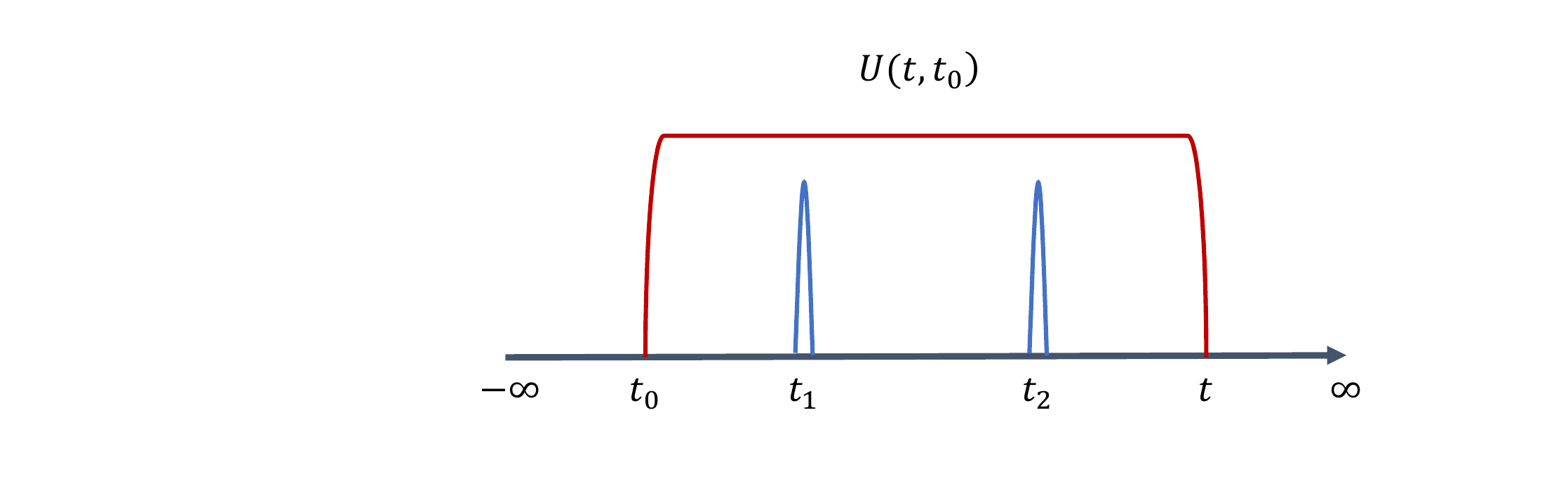}
\caption{The two-photon input is modelled as two separated $\delta$ pulses. The first photon enters and interacts with the system at $t=t_1$, then the second photon enters at $t=t_2$. The unitary operator $U(t,t_0)$ governs the entire process.}
\label{fig1}
\end{figure}

Next, we will define a continuous-mode two-photon state. Heuristically, we consider a singular input state $b_i^\dagger(t_1)b_j^\dagger(t_2)|0\rangle$ (See Figure~\ref{fig1}) which contains two impulses of single-photon inputs at $t_1$ and $t_2$, respectively. When the initial state of the system is $|0_s\rangle$, the unitary evolution of the system-field state $|\Psi(t)\rangle$ is given by
\begin{equation}
|\Psi(t)\rangle=U(t,t_0)b_i^\dagger(t_1)b_j^\dagger(t_2)|00_s\rangle.
\end{equation}
Here $t_1,t_2\in[t_0,t]$ is required in order to guarantee that $U(t,t_0)$ covers the effective interaction process and so $|\Psi(t)\rangle$ is the output state. Since the continuous-mode single-photon pulse can be modelled as the superposition of single-photon impulses \cite{Gough12}, the general form of an uncorrelated two-photon input state can be defined as
\begin{equation}\label{genin}
|1_{\xi_1}1_{\xi_2}\rangle=\int_{-\infty}^\infty dt_1\int_{-\infty}^\infty dt_2b^\dagger(t_1)\xi_1(t_1)b^\dagger(t_2)\xi_2(t_2)|0\rangle,
\end{equation}
where each $\xi_q(\cdot)=[\xi_{q1},\cdot\cdot\cdot,\xi_{qK}]^T(\cdot),q=1,2$ is the collection of pulse functions over the $K$ channels for each single photon. The pulse functions satisfy the normalization condition $\sum_{k=1}^K\int_{-\infty}^\infty|\xi_{qk}(t)|^2dt=1$. According to the generalized definition Eq.~(\ref{genin}), each single-photon input could be superposed over the $K$ channels. Since the input state Eq.~(\ref{genin}) is defined on $(-\infty,+\infty)$, we must let $t\rightarrow\infty$ and $t_0\rightarrow-\infty$ in order to obtain the correct output state. The joint (system plus field) output state can thus be calculated by
\begin{eqnarray}\label{ustate}
|\Psi(\infty)\rangle&=&U(\infty,-\infty)|1_{\xi_1}1_{\xi_2}0_s\rangle\nonumber\\
&=&\int_{-\infty}^\infty dt_1\int_{-\infty}^\infty dt_2U(\infty,-\infty)b^\dagger(t_1)\xi_1(t_1)b^\dagger(t_2)\xi_2(t_2)|00_s\rangle.
\end{eqnarray}
Here $U(\infty,-\infty)$ is well-defined due to Eq.~(\ref{inout}). We need to convert Eq.~(\ref{ustate}) to a computable form. Note that the output state $|\Psi(\infty)\rangle$ is a two-photon state with two excitations in the field, according to Assumption \ref{assumption1}. Here we have ignored the component $b_i^\dag(\tau)|01_s\rangle,\ \tau\in(-\infty,\infty)$ in the output state, where $|1_s\rangle$ is the system state containing one excitation. This component vanishes in the steady-state limit by taking $t_0\rightarrow-\infty$. More details on the steady-state limit can be found in the next section. Hence, the basis vectors of $|\Psi(\infty)\rangle$ are
\begin{equation}\label{bavec}
b_i^\dag(\tau_1)b_j^\dag(\tau_2)|00_s\rangle,\tau_1,\tau_2\in(-\infty,\infty).
\end{equation}
As a result, the output field state is calculated by
\begin{eqnarray}\label{rhof}
\fl |\Psi_{field}(\infty)\rangle&=&\langle 0_s|\int_{-\infty}^\infty dt_1\int_{-\infty}^\infty dt_2U(\infty,-\infty)b^\dagger(t_1)\xi_1(t_1)b^\dagger(t_2)\xi_2(t_2)|00_s\rangle.
\end{eqnarray}
Inserting the identity, i.e. $\sum_{i,j=1}^K\int_{-\infty}^\infty\int_{-\infty}^\infty d\tau_1d\tau_2b_i^\dag(\tau_1)b_j^\dag(\tau_2)|00_s\rangle\langle00_s|b_i(\tau_1)b_j(\tau_2)$ onto the two-photon subspace, we obtain
\begin{equation}
|\Psi_{field}(\infty)\rangle=\sum_{i,j=1}^K\int_{-\infty}^\infty\int_{-\infty}^\infty d\tau_1d\tau_2\xi^{'}_{ij}(\tau_1,\tau_2)b_i^\dag(\tau_1)b_j^{\dag}(\tau_2)|0\rangle,
\end{equation}
with the coefficients $\{\xi^{'}_{ij}(\tau_1,\tau_2)\}$ of the basis vectors defined by
\begin{eqnarray}
\fl \xi^{'}_{ij}(\tau_1,\tau_2)=\langle00_s|b_j(\tau_2)b_i(\tau_1)\int_{-\infty}^\infty dt_1\int_{-\infty}^\infty dt_2U(\infty,-\infty)b^\dagger(t_1)\xi_1(t_1)b^\dagger(t_2)\xi_2(t_2)|00_s\rangle.\label{rhor1}
\end{eqnarray}
According to this expression, $|\xi^{'}_{ij}(\tau_1,\tau_2)|^2d\tau_1d\tau_2$ is the probability of emitting the two photons to the $i$-th and $j$-th channels of the output field during $[\tau_1,\tau_1+d\tau_1)$ and $[\tau_2,\tau_2+d\tau_2)$, respectively. The output field state $|\Psi_{field}(\infty)\rangle$ is analytically computable if $\{\xi^{'}_{ij}(\tau_1,\tau_2)\}$ are analytically computable. Inserting the identity $U(\infty,-\infty)U^\dag(\infty,-\infty)$ helps simplify the expression, which will enable the analytical computability of $\xi^{'}_{ij}(\tau_1,\tau_2)$ to be studied by
\begin{eqnarray}\label{fdag}
\fl \xi^{'}_{ij}(\tau_1,\tau_2)=\langle00_s|\int_{-\infty}^\infty dt_1\int_{-\infty}^\infty dt_2U(\infty,-\infty)U^\dag(\infty,-\infty)b_j(\tau_2)U(\infty,-\infty)\nonumber\\
\fl \times U^\dag(\infty,-\infty)b_i(\tau_1)U(\infty,-\infty)b^\dagger(t_1)\xi_1(t_1)b^\dagger(t_2)\xi_2(t_2)|00_s\rangle\nonumber\\
\fl =\langle00_s|\int_{-\infty}^\infty dt_1\int_{-\infty}^\infty dt_2b_{j,out}(\tau_2)b_{i,out}(\tau_1)b^\dagger(t_1)\xi_1(t_1)b^\dagger(t_2)\xi_2(t_2)|00_s\rangle,
\end{eqnarray}
where $b_{i,out}(\cdot),b_{j,out}(\cdot)$ are the $i$-th and $j$-th components of $b_{out}(\cdot)$, respectively. Here we have used the property Eq.~(\ref{pas2}) and the input-output relation Eq.~(\ref{inout}). Note that the pulse function $\xi^{'}_{ij}(\tau_1,\tau_2)$ is symmetric with respect to $\tau_1=\tau_2$, which is due to the indistinguishability of photons. Making use of Eq.~(\ref{nopr2}), Eq.~(\ref{fdag}) can be decomposed into four terms
\begin{equation}\label{fdag1}
\xi^{'}_{ij}(\tau_1,\tau_2)=\int_{-\infty}^\infty dt_1\int_{-\infty}^\infty dt_2[\tilde{\mathcal O}_I+\tilde{\mathcal O}_{II}+\tilde{\mathcal O}_{III}+\tilde{\mathcal O}_{IV}],
\end{equation}
with
\begin{eqnarray}\label{fdag2}
\tilde{\mathcal O}_I&=&\langle00_s|b_{j}(\tau_2)b_{i}(\tau_1)b^\dagger(t_1)\xi_1(t_1)b^\dagger(t_2)\xi_2(t_2)|00_s\rangle,\nonumber\\
\tilde{\mathcal O}_{II}&=&\langle00_s|L_{j}(\tau_2)b_{i}(\tau_1)b^\dagger(t_1)\xi_1(t_1)b^\dagger(t_2)\xi_2(t_2)|00_s\rangle,\nonumber\\
\tilde{\mathcal O}_{III}&=&\langle00_s|b_{j}(\tau_2)L_{i}(\tau_1)b^\dagger(t_1)\xi_1(t_1)b^\dagger(t_2)\xi_2(t_2)|00_s\rangle,\nonumber\\
\tilde{\mathcal O}_{IV}&=&\langle00_s|L_{j}(\tau_2)L_{i}(\tau_1)b^\dagger(t_1)\xi_1(t_1)b^\dagger(t_2)\xi_2(t_2)|00_s\rangle.
\end{eqnarray}
Thus far, we have developed the formalism for the calculation of the response of a general quantum system to two-photon input. The analytical expression of the two-photon output state can be obtained if Eq.~({\ref{fdag2}}) can be calculated. Eq.~({\ref{fdag2}}) only contains the singular field operators and the Heisenberg-picture system operators which can be solved using their QSDEs.

Interestingly, the four terms in Eq.~({\ref{fdag2}}) are not the most general form for the decomposition Eq.~(\ref{fdag1}). We are able to further decompose $\{L_i\}$ as
\begin{equation}\label{fdag3}
L_i=\theta_i^TL^{'},\quad L^{'}=[L_1^{'}\cdot\cdot\cdot L_M^{'}]^T,\ i=1,2,\cdot\cdot\cdot,K,
\end{equation}
with $\{\theta_i\}$ being constant column vectors. In other words, the coupling operator $L_i$ can be written as linear combination of a set of component operators $\{L_m^{'},m=1,2,\cdot\cdot\cdot,M\}$. Note that $L_i=L_i^{'}$ is a special case for Eq.~(\ref{fdag3}) but $K$ and $M$ may not be the same in general. Using Eq.~(\ref{fdag3}), we can re-express Eq.~(\ref{fdag2}) as the linear combination of the following four terms
\begin{eqnarray}\label{fdag4}
\mathcal O_I&=&\langle00_s|b_{j}(\tau_2)b_{i}(\tau_1)b^\dagger(t_1)\xi_1(t_1)b^\dagger(t_2)\xi_2(t_2)|00_s\rangle,\nonumber\\
\mathcal O_{II}&=&\langle00_s|L_{n}^{'}(\tau_2)b_{i}(\tau_1)b^\dagger(t_1)\xi_1(t_1)b^\dagger(t_2)\xi_2(t_2)|00_s\rangle,\nonumber\\
\mathcal O_{III}&=&\langle00_s|b_{j}(\tau_2)L_{m}^{'}(\tau_1)b^\dagger(t_1)\xi_1(t_1)b^\dagger(t_2)\xi_2(t_2)|00_s\rangle,\nonumber\\
\mathcal O_{IV}&=&\langle00_s|L_{n}^{'}(\tau_2)L_{m}^{'}(\tau_1)b^\dagger(t_1)\xi_1(t_1)b^\dagger(t_2)\xi_2(t_2)|00_s\rangle
\end{eqnarray}
for $n=1,2,\cdot\cdot\cdot,M$. In most cases, it is more convenient to deal with Eq.~(\ref{fdag4}) rather than Eq.~(\ref{fdag2}). This is because the information of the linear combination (e.g. coefficients of the component operators) does not affect the analytical computability of the output state. As a consequence, component operators normally have simpler forms compared to $\{L_i\}$, and so applying QSDE analysis on component operators often leads to simpler conditions. As we will show in Section \ref{sectl}, the coupling operator for a two-level atom could be $L=\sqrt{\kappa}\sigma_-,\kappa>0$. However, the QSDE analysis of $L^{'}=\sigma_-$ is sufficient for proving the analytical computability of the output state.

The response to a general two-photon input state can be analyzed using the same formalism, simply by replacing the uncorrelated state Eq.~(\ref{genin}) with the general two-photon state in the derivations. For example, if the system couples to the environment via a single channel, a general two-photon input is written as
\begin{equation}\label{generaltwo}
|2_\xi\rangle=\int_{-\infty}^\infty dt_1\int_{-\infty}^\infty dt_2\xi(t_1,t_2)b^\dagger(t_1)b^\dagger(t_2)|0\rangle,
\end{equation}
with the normalization condition $\int_{-\infty}^\infty dt_1\int_{-\infty}^\infty dt_2|\xi(t_1,t_2)|^2dt_1dt_2=1$. It is easy to see that the derivations remain the same if we use the general input state Eq.~(\ref{generaltwo}) instead.

\section{Analytical computability of the output field state}\label{AC}
In this section we focus on the analytical computability of Eq.~(\ref{fdag4}). It is easy to see that $\mathcal O_I$ can be exactly calculated using the commutation relation of the field operators $\{b_i,b_i^\dag\}$. The calculation of the rest terms in Eq.~(\ref{fdag4}) relies on the Heisenberg-picture dynamics of the operator $L_m^{'}(t)$, which is characterized by the QSDE
\begin{equation}\label{one}
\dot{L}_m^{'}(t)=\mathcal L^\dagger(L_m^{'}(t))+[L^\dag(t),L_m^{'}(t)]b(t)+b^\dag(t)[L_m^{'}(t),L(t)],
\end{equation}
where $[L^\dag(t),L_m^{'}(t)]=[[L_1^\dag(t),L_m^{'}(t)],\cdot\cdot\cdot,[L_K^\dag(t),L_m^{'}(t)]]$ is a row vector of commutators. By Eq.~(\ref{one}), we can write the QSDE of $L^{'}(t)$ in a vector form
\begin{equation}\label{node}
\dot{L}^{'}(t)=\mathcal L^\dagger(L^{'}(t))+\tilde{B}(t)b(t)+b^\dag(t)\tilde{C}(t),
\end{equation}
where $\mathcal L^\dagger(L^{'}(t))=[\mathcal L^\dagger(L_1^{'}(t))\ \cdot\cdot\cdot\ \mathcal L^\dagger(L_M^{'}(t))]^T$. The operator matrix $\tilde{B}$ is defined by $\tilde{B}=\{[L_j^\dag,L_m^{'}]\}$, i.e. the $(m,j)$-th entry of $\tilde{B}$ is the commutator $[L_j^\dag,L_m^{'}]$. Similarly, the operator matrix $\tilde{C}$ is defined by $\tilde{C}=\{[L_m^{'},L_j]\}$, i.e. the $(m,j)$-th entry of $\tilde{C}$ is the commutator $[L_m^{'},L_j]$. Eq.~(\ref{node}) implies
\begin{equation}\label{ode1}
\langle00_s|\dot{L}^{'}(t)=\langle00_s|(\mathcal L^\dagger(L^{'}(t))+\tilde{B}(t)b(t)).
\end{equation}
Therefore, if the following conditions
\begin{eqnarray}
&&\mathcal L^\dagger(L^{'}(t))=AL^{'}(t),\label{as1}\\
&&\langle00_s|\tilde{B}=\langle00_s|B,\label{as2}
\end{eqnarray}
hold with constant matrices $A$ and $B$, then Eq.~(\ref{ode1}) is a solvable ordinary differential equation (ODE). By Eq.~(\ref{as1}), each $\mathcal L^\dagger(L_m^{'}(t))$ is required to be a linear combination of the component operators. Furthermore, it is required that $A$ must be Hurwitz. Hurwitz means that the real parts of the eigenvalues of $A$ are strictly negative. A Hurwitz matrix $A$ can remove the instantaneous response of the system and keep only the steady-state dynamics. In stability theory, $A$ being Hurwitz is equivalent to the asymptotic stability of the linear system Eq.~(\ref{as1}). In most cases, since the energy is conserved in the overall system (see Assumption \ref{assumption1}), the photons will eventually leak to the fields and the system will be stabilized to its ground state. As a result, $A$ being Hurwitz could be a natural property of the systems considered in this paper.

According to Eq.~(\ref{as2}), the elements of $B=\{b_{mj}\}$ are given by $\langle00_s|[L_j^\dag,L^{'}_m]=\langle00_s|b_{mj}$. In other words, $|00_s\rangle$ is an eigenvector of the commutators $\{[(L_m^{'})^\dag,L_j]\}$. Solving Eq.~(\ref{ode1}) and letting $t_0\rightarrow-\infty$ we have
\begin{equation}\label{lsol}
\langle00_s|L^{'}(t)=\langle00_s|\int_{-\infty}^te^{A(t-r)}Bb(r)dr,
\end{equation}
which is in the form of a convolution. Obviously, a frequency-domain relation naturally follows from the convolution. Here we have used the condition $A$ being Hurwitz, so that the instantaneous term in the solution of the ODE converges to zero as $t_0\rightarrow-\infty$. It is worth mentioning that the steady-state solution Eq.~(\ref{lsol}) only contains excitations in the field, which is consistent with our discussion in the last section.

Alternatively, we can write Eq.~(\ref{lsol}) as
\begin{equation}\label{lsol1}
\langle00_s|L_m^{'}(t)=\langle00_s|\int_{-\infty}^tg_m(t-r)b(r)dr,
\end{equation}
where $g_m(t-r)$ is the $m$-th column of $e^{A(t-r)}B$. Using Eq.~(\ref{lsol1}), we can write $\mathcal O_{II}$ as
\begin{equation}
\mathcal O_{II}=\langle00_s|\int_{-\infty}^{\tau_2}g_n(\tau_2-r)b(r)drb_{i}(\tau_1)b^\dagger(t_1)\xi_1(t_1)b^\dagger(t_2)\xi_2(t_2)|00_s\rangle,
\end{equation}
which can be readily calculated using the commutation relations. Similarly, we can write $\mathcal O_{IV}$ as
\begin{equation}\label{iv1}\label{end}
\mathcal O_{IV}=\int_{-\infty}^{\tau_2}g_n(\tau_2-r)dr\langle00_s|b(r)L^{'}_{m}(\tau_1)b^\dagger(t_1)\xi_1(t_1)b^\dagger(t_2)\xi_2(t_2)|00_s\rangle.
\end{equation}
By Eq.~(\ref{iv1}), $\mathcal O_{III}$ and $\mathcal O_{IV}$ are analytically computable if
\begin{equation}\label{iv2}
\langle00_s|b_j(\tau)L^{'}_{m}(\tau_1)b^\dagger(t_1)\xi_1(t_1)b^\dagger(t_2)\xi_2(t_2)|00_s\rangle
\end{equation}
is analytically computable for arbitrary $j$ and $\tau$. Now the question is whether we can express Eq.~(\ref{iv2}) using field operators only so that we can again apply the commutation relations.

The formal integration of Eq.~(\ref{node}) leads to
\begin{equation}\label{fode}
L^{'}(t)=\int_{-\infty}^te^{A(t-r)}(\tilde{B}(r)b(r)+b^\dag(r)\tilde{C}(r))dr.
\end{equation}
Substituting Eq.~(\ref{fode}) into Eq.~(\ref{iv2}), and again making use of the commutation relation, it is straightforward to show that the sufficient condition for Eq.~(\ref{iv2}) to be analytically computable is that the following terms
\begin{eqnarray}\label{term1}
\fl \langle00_s|b_{j}(\tau)[(L_n^{'})^\dag(r),L_m^{'}(r)]b_i^\dag(t_q)|00_s\rangle,\quad \langle00_s|[L_n^{'}(r),L_m^{'}(r)]b_i^\dag(t_1)b_j^\dag(t_2)|00_s\rangle
\end{eqnarray}
are analytically computable for arbitrary $i,j,m,n$. The analytical computability of the two terms is largely dependent on the commutation property of the component operators. We consider two typical cases:

\subsection{$[(L_n^{'})^\dag,L_m^{'}]=$constant and $[L_n^{'},L_m^{'}]=$constant}
In this case, Eq.~(\ref{term1}) can be readily computed using the commutation relation. According to the above discussion, the output state is analytically computable. Systems that satisfy this condition will be discussed in Section~\ref{secpl}-\ref{secop}.

\subsection{$[(L_n^{'})^\dag,L_m^{'}]$ is a nontrivial operator and $[L_n^{'},L_m^{'}]=$constant}
Since $[L_n^{'},L_m^{'}]$ is a constant, we have $\langle00_s|[L_n^{'}(r),L_m^{'}(r)]b_i^\dag(t_1)b_j^\dag(t_2)|00_s\rangle=0$. When $[(L_n^{'})^\dag,L_m^{'}]$ is a nontrivial operator, we can perform integration on Eq.~(\ref{nopr1}) to obtain
\begin{eqnarray}\label{iv4}
&&\langle00_s|b_{j}(\tau)[(L_n^{'})^\dag(r),L_m^{'}(r)]b^\dag_{i}(t_q)|00_s\rangle\nonumber\\
&=&\langle00_s|b_{j}(\tau)\int_{-\infty}^rdse^{c_{mn}(r-s)}\{[L^\dag(s),[(L_n^{'})^\dag(s),L_m^{'}(s)]]b(s)\nonumber\\
&+&b^\dag(s)[[(L_n^{'})^\dag(s),L_m^{'}(s)],L(s)]\}b^\dag_{i}(t_q)|00_s\rangle\nonumber\\
&=&\delta(s-t_q)\langle00_s|b_{j}(\tau)[\int_{-\infty}^rdse^{c_{mn}(r-s)}[L_{i}^\dag(s),[(L_n^{'})^\dag(s),L_m^{'}(s)]]|00_s\rangle\nonumber\\
&+&\delta(s-\tau)\langle00_s|[\int_{-\infty}^rdse^{c_{mn}(r-s)}[[(L_n^{'})^\dag(s),L_m^{'}(s)],L_{j}(s)]b^\dag_{i}(t_q)|00_s\rangle
\end{eqnarray}
if the condition
\begin{equation}
\mathcal L^\dagger([(L_n^{'})^\dag,L_m^{'}])=c_{mn}[(L_n^{'})^\dag,L_m^{'}],\ \Re(c_{mn})<0
\end{equation}
is satisfied. Now recall Eq.~(\ref{lsol1}), where we have proven that the component operators sandwiched between single photon state and vacuum state are exactly calculable. As a result, Eq.~(\ref{iv4}) is analytically computable if
\begin{equation}\label{kij}
[[(L_n^{'})^\dag,L_m^{'}],L]=\Lambda L^{'},
\end{equation}
with $\Lambda$ being a $K\times M$ constant matrix.

We have derived a set of algebraic conditions which are easily checkable. The component operators $\{L_m^{'}\}$ in these conditions may not necessarily be the actual coupling operators between the system and the fields. Nevertheless, the actual coupling operators must be linear combinations of the component operators. Consequently, for the purpose of computing the two-photon output state, the key thing is to identify a set of component operators $\{L_m^{'}\}$ which satisfy the sufficient conditions. In the subsequent sections, we will show that the exact form of the time-domain output states can be obtained for the systems whose modelling and time-domain calculation are difficult using conventional approaches. Particularly, we obtain the exact real-time two-photon output state for one-channel and two-channel scattering by a two-level emitter. The scattering of a two-level emitter is a problem of critical importance and has been extensively studied in the literature.

\section{Passive linear network}\label{secpl}
We consider a linear network composed of single-mode harmonic oscillators which are coupled together via interconnection. This may refer to an optical network. An optical linear network can be used to process the information encoded in photons \cite{Adami99,Knill01}. Due to the application in linear quantum computation and photonic circuitry, the linear optical network has been extensively studied in the engineering community \cite{Zhang12,Tezak5270,Hendra09,Gough2012}. It is well known that there are two basic types of interconnections for the construction of a network: cascade and direct interaction. Coherent feedback and other types of interconnections can be built up from the two basic types. Therefore, we just need to prove the analytical computability of the response of cascaded and directly coupled linear systems.

\begin{figure}
\includegraphics[scale=0.5]{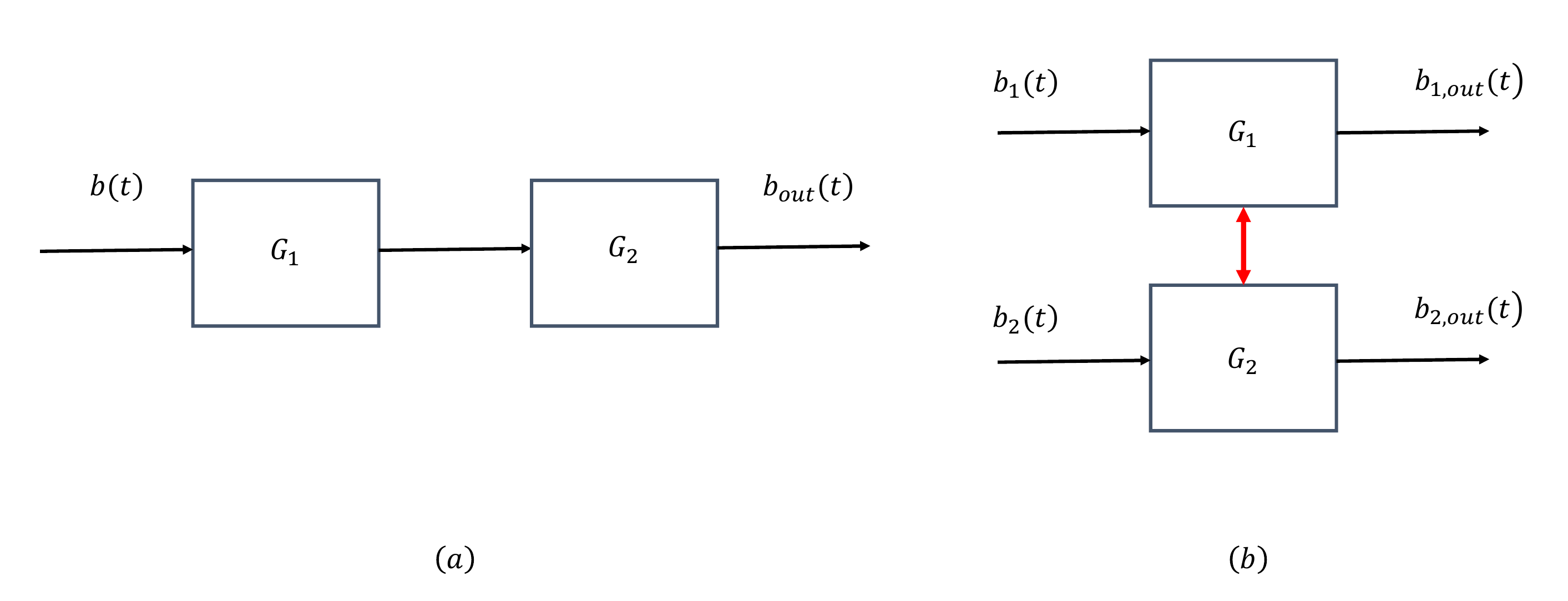}
\caption{(a) The cascade interconnection of $G_1$ and $G_2$. The cascaded system has single input and output channel. (b) There is a linear interaction between $G_1$ and $G_2$. The combined system has two input and two output channels.}
\label{fig4}
\end{figure}

A cascade connection of two single-mode open cavities with internal modes $a_1,a_2$ and resonant frequencies $\omega_1,\omega_2$ is depicted in Figure~\ref{fig4}(a). Suppose the coupling operators for the cavities are $L_1=\sqrt{\kappa_1}a_1$ and $L_2=\sqrt{\kappa_2}a_2$, $\kappa_1,\kappa_2>0$, respectively. $\kappa_1,\kappa_2$ are the decay rates of the cavities. The output signal of the first cavity is fed into the second cavity as the input. Employing the network theory, the cascaded system is described by the coupling operator $L=\sqrt{\kappa_1}a_1+\sqrt{\kappa_2}a_2$ and Hamiltonian $H_0=\frac{\omega_1}{2}a_1^\dag a_1+\frac{\omega_2}{2}a_2^\dag a_2+\frac{\sqrt{\kappa_1\kappa_2}}{2\mbox{i}}(a_2^\dag a_1-a_1^\dag a_2)$ \cite{PhysRevLett.70.2269,PhysRevLett.70.2273,Gough09,James10}. $L$ is a linear combination of $a_1$ and $a_2$ and so we can let $L^{'}=[a_1\quad a_2]^T$. By Eq.~(\ref{nopr1}) we have the linear equations
\begin{eqnarray}
\mathcal L^\dagger({a}_1)=-\frac{1}{2}(\mbox{i}\omega_1+\kappa_1)a_1+\frac{\sqrt{\kappa_1\kappa_2}}{2}a_2,\nonumber\\
\mathcal L^\dagger({a}_2)=-\frac{1}{2}(\mbox{i}\omega_2+\kappa_2)a_2-\frac{\sqrt{\kappa_1\kappa_2}}{2}a_1,
\end{eqnarray}
whose vector form is
\begin{equation}
\mathcal L^\dagger(L^{'})=AL^{'},\quad A=\left(
\begin{array}{cc}
-\frac{1}{2}(\mbox{i}\omega_1+\kappa_1)&\frac{\sqrt{\kappa_1\kappa_2}}{2}\\
-\frac{\sqrt{\kappa_1\kappa_2}}{2}&-\frac{1}{2}(\mbox{i}\omega_2+\kappa_2)
\end{array}
\right).
\end{equation}
$A$ is Hurwitz if the real part of $A$ is negative definite. Since
\begin{equation}
\Re(A)=\left(
\begin{array}{cc}
-\frac{1}{2}\kappa_1&\frac{\sqrt{\kappa_1\kappa_2}}{2}\\
-\frac{\sqrt{\kappa_1\kappa_2}}{2}&-\frac{1}{2}\kappa_2
\end{array}
\right)<0
\end{equation}
holds for any $\kappa_1,\kappa_2>0$, we can conclude that $A$ is Hurwitz. Furthermore, since the commutators $[a_1^\dag,a_1]=-1,[a_1,a_1]=0,[a_2^\dag,a_2]=-1,[a_2,a_2]=0$ are constants, the two-photon output state can be exactly calculated.

The directly coupled system as plotted in Figure~\ref{fig4}(b) is described by the linear coupling operator $L=[\sqrt{\kappa_1}a_1\quad \sqrt{\kappa_2}a_2]^T$ and Hamiltonian $H_0=\frac{\omega_1}{2}a_1^\dag a_1+\frac{\omega_2}{2}a_2^\dag a_2+\gamma(a_2^\dag a_1+a_1^\dag a_2)$ \cite{Gough09,James10}. $\gamma$ is the coupling strength. The two subsystems are coupled via a linear interaction term $\gamma(a_2^\dag a_1+a_1^\dag a_2)$.  In this case, we can still let $L^{'}=[a_1\quad a_2]^T$ as in the cascade case. We can obtain the linear equations
\begin{eqnarray}
\mathcal L^\dagger({a}_1)=-\frac{1}{2}(\mbox{i}\omega_1+\kappa_1)a_1-\mbox{i}\gamma a_2,\nonumber\\
\mathcal L^\dagger({a}_2)=-\frac{1}{2}(\mbox{i}\omega_2+\kappa_2)a_2-\mbox{i}\gamma a_1.
\end{eqnarray}
The coefficient matrix of $\mathcal L^\dagger(L^{'})=AL^{'}$ is
\begin{equation}
A=\left(
\begin{array}{cc}
-\frac{1}{2}(\mbox{i}\omega_1+\kappa_1)&-\mbox{i}\gamma\\
-\mbox{i}\gamma&-\frac{1}{2}(\mbox{i}\omega_2+\kappa_2)
\end{array}
\right)
\end{equation}
which is Hurwitz. Therefore, the output state is exactly calculable for the directly coupled system.

\section{Optomechanical system}\label{secop}
\begin{figure}
\includegraphics[scale=0.5]{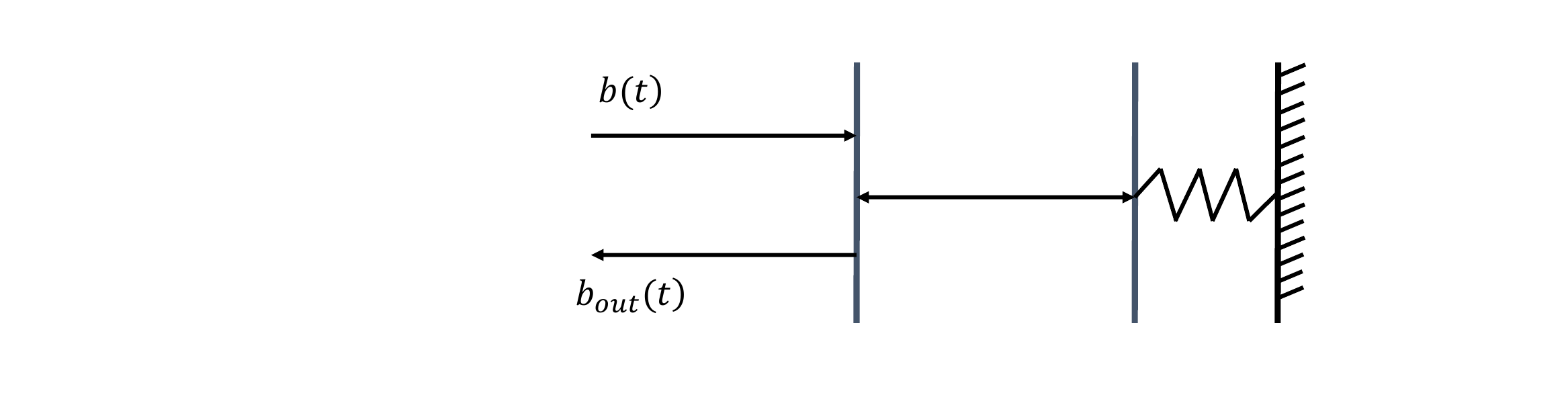}
\caption{The optomechanical system has single input and output channel. The decay of the mechanical mode is negligible compared to the decay of the cavity mode.}
\label{opto}
\end{figure}

An optomechanical system may follow linearized dynamical equation under certain circumstances \cite{RevModPhys.86.1391,Naoki14}. The optomechanical system as shown in Figure~\ref{opto} is composed of a linear cavity and a mechanical oscillator in interaction. The linearized system Hamiltonian is given by $H_0=\frac{\omega_c}{2}c^\dag c+\frac{\omega_m}{2}a^\dag a+\gamma(c^\dag+c)(a^\dag+a)$ \cite{RevModPhys.86.1391}, where $\gamma$ can be made a real number. $c$ is the cavity mode and $a$ is the mechanical mode. The optomechanical system couples to the external field by the coupling operator $L=\sqrt{\kappa}c$. Letting $L_1^{'}=c$, we can derive
\begin{equation}
\mathcal L^\dagger(c)=-\frac{1}{2}(\mbox{i}\omega_c+\kappa)c-\mbox{i}\gamma(a+a^\dag).
\end{equation}
Although the coupling operator $L$ contains $c$ only, the generator of $c$ is dependent on $a,a^\dag$. This observation motivates us to define $L^{'}=[c\quad c^\dag\quad a\quad a^\dag]^T$. Note that $L$ is still a linear combination of the component operators, with the coefficients on $c^\dag,a,a^\dag$ being zero. Since
\begin{equation}
\mathcal L^\dagger(a)=-\frac{1}{2}\mbox{i}\omega_ma-\mbox{i}\gamma(c+c^\dag),
\end{equation}
we can express $\mathcal L^\dagger(L^{'})$ as a linear equation
\begin{equation}
\mathcal L^\dagger(L^{'})=AL^{'}=\left(
\begin{array}{cccc}
-\frac{1}{2}(\mbox{i}\omega_c+\kappa)&0&-\mbox{i}\gamma&-\mbox{i}\gamma\\
0&\frac{1}{2}(\mbox{i}\omega_c-\kappa)&-\mbox{i}\gamma&-\mbox{i}\gamma\\
-\mbox{i}\gamma&-\mbox{i}\gamma&-\frac{1}{2}\mbox{i}\omega_m&0\\
\mbox{i}\gamma&\mbox{i}\gamma&0&\frac{1}{2}\mbox{i}\omega_m
\end{array}
\right)L^{'}.
\end{equation}
$A$ is Hurwitz due to $\Re(A)<0$, which is also a consequence of the passivity of the system. Additionally, the commutations between $c,c^\dag,a,a^\dag$ are all constants. Thus the response of the optomechanical system to two-photon input can be exactly computed.

\section{Two-level emitter}\label{sectl}
We consider a two-level emitter in interaction with the photons propagating in the waveguide. The coupling of the two-level emitter to the optical fields is often modelled by $\sum_i\sqrt{\kappa_i}(\sigma_-b_i^\dag(\omega)+b_i(\omega)\sigma_+)$, where $b_i(\omega)$ is the annihilation operator for the $i$-th mode of the field. When the photons propagate along the waveguide unidirectionally, there is only one coupling channel. If there are two modes for the travelling photons (e.g. left-propagating and right-propagating), we should model the interaction using two coupling channels. To be specific, the system is coupled to the left-going mode via one channel, and coupled to the right-going mode via another channel. Since the interaction is energy-preserving, the system is passive when there exists no additional control that pumps energy into the system. We apply the results of Sec.~\ref{AC} to two processes of wide interests: stimulated emission of a two-level atom and the inelastic scattering of two photons.

\subsection{Simulation of stimulated emission}
The stimulated emission can be modelled by a two-level emitter interacting with a single-channel input field via $L=\sqrt{\kappa_1}\sigma_-$. Without loss of generality we assume the system Hamiltonian is $H_0=0$. In particular, introducing a free Hamiltonian of the form $H_0=\frac{\omega_c}{2}\sigma_z$ will only induce an additional harmonic component with frequency $\omega_c$ in the pulse functions.

When the system is fully excited and coupled to vacuum, the spontaneous emission rate is $\kappa_1$. However, if a second incoming photon interacts with the population-inverted emitter, the emission of a photon may either accelerate or slow down, depending on the exact form of the input pulse.

It is easy to verify the commutation relations of $L^{'}=\sigma_-$ by $[\sigma_+,\sigma_-]=\sigma_z,[\sigma_+,[\sigma_+,\sigma_-]]=-2\sigma_+$ and $\langle00_s|[\sigma_+,\sigma_-]=\langle00_s|\sigma_z=-\langle00_s|$. Also, we can obtain\
\begin{eqnarray}
\dot{\sigma}_-(t)&=&-\frac{\kappa_1}{2}\sigma_-(t)+\sqrt{\kappa_1}\sigma_z(t)b(t)\nonumber\\
\dot{\sigma}_z(t)&=&-\kappa_1(I+\sigma_z(t))-2\sqrt{\kappa_1}(b^\dag(t)\sigma_-(t)+\sigma_+(t)b(t)).
\end{eqnarray}
Therefore, we can conclude $\mathcal L^\dagger(\sigma_-)=-\frac{\kappa_1}{2}\sigma_-$ and $\mathcal L^\dagger([\sigma_+,\sigma_-])=\mathcal L^\dagger(\sigma_z)=-\kappa_1(I+\sigma_z)$.

Noting that the output field state is exactly solvable, we just need to design the pulse shapes so that the two-photon interaction could simulate the stimulated emission process. Intuitively, the first photon, which is followed immediately by the second photon, should be able to fully excite the system from the ground state to the excited state. For this reason, we choose the pulse function of the first photon to be the following form
\begin{equation}\label{sefp}
\xi_1(t_1)=-\sqrt{\gamma}e^{\frac{\gamma}{2}t_1}(1-u(t_1)),
\end{equation}
where $u(t_1)$ is the Heaviside step function and $\gamma$ is a controllable parameter. Eq.~(\ref{sefp}) is the famous rising exponential pulse which can perfectly transfer the single photon to the two-level system at $t=0$ \cite{Cirac97,Naoki2014} when we let $\gamma=\kappa_1$. As a result, the second incoming photon should be defined on $t_2\in(0,+\infty)$. Following the procedures in Sec.~\ref{AC}, we can exactly calculate the output field state to be
\begin{equation}
|\Psi_{field}(\infty)\rangle=\int_{-\infty}^\infty\int_{-\infty}^\infty d\tau_1d\tau_2\xi^{'}(\tau_1,\tau_2)b^\dag(\tau_1)b^\dag(\tau_2)|00_s\rangle,
\end{equation}
with the analytical form of the pulse function given by
\begin{eqnarray}
\xi^{'}(\tau_1,\tau_2)&=&\sqrt{\kappa_1}e^{-\frac{\kappa_1}{2}\tau_1}\xi_2(\tau_2)+\sqrt{\kappa_1}e^{-\frac{\kappa_1}{2}\tau_2}\xi_2(\tau_1)\nonumber\\
&+&\kappa_1\int_{\tau_1}^{\tau_2}e^{-\frac{\kappa_1}{2}(\tau_1+\tau_2-s)}\xi_2(s)ds,\ \tau_2\geq\tau_1\geq0.
\end{eqnarray}
\begin{figure}
\includegraphics[scale=0.8]{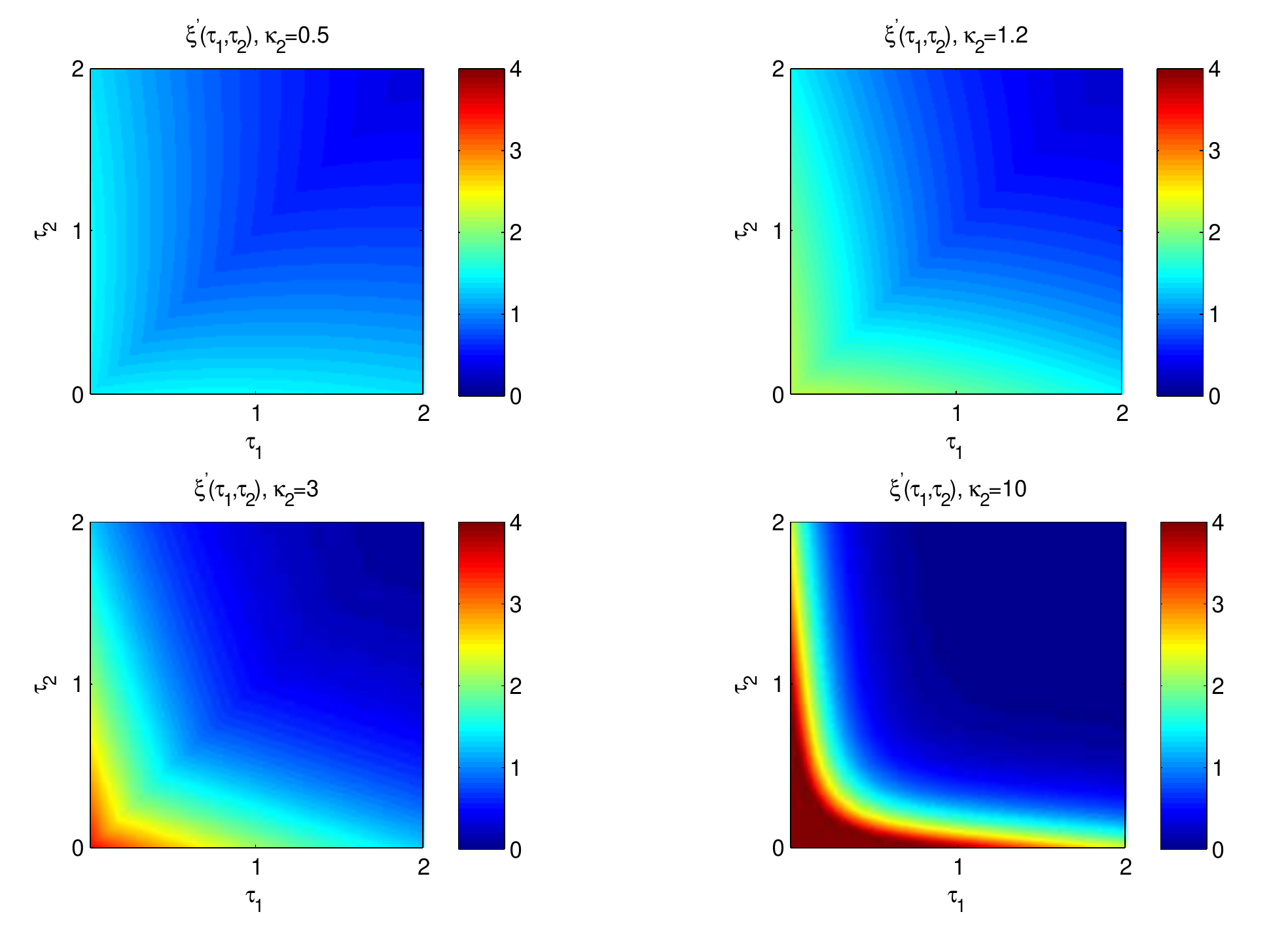}
\caption{ The value of $\xi^{'}(\tau_1,\tau_2)$ for different $\kappa_2$, calculated using Eq.~(\ref{ste}). The result is symmetric with respect to $\tau_1=\tau_2$. $\kappa_1=1$. $\kappa_2=3$ (left, lower row) gives the optimal performance. When $\kappa_2=0.5$ or $\kappa_2=1.2$, the emitter decays slowly. When $\kappa_2=10$, there is a high probability that the interval between the emission of the two photons is longer than $1$. }
\label{fig2}
\end{figure}
It is straightforward to employ this analytical form to study the optimal stimulated emission, which could be used for photon amplification \cite{Valente12}. For example, if the pulse function of the second photon is defined as $\xi_2(t_2)=\sqrt{\kappa_2}e^{-\frac{\kappa_2}{2}t_2}u(t_2)$ with $\kappa_2$ being a controllable parameter, then we have
\begin{equation}\label{ste}
\fl \xi^{'}(\tau_1,\tau_2)=\sqrt{\kappa_1\kappa_2}e^{-\frac{\kappa_1+\kappa_2}{2}\tau_1}\{(1+\frac{2\sqrt{\kappa_1}}{\kappa_1-\kappa_2})e^{-\frac{\kappa_1}{2}(\tau_2-\tau_1)}+(1-\frac{2\sqrt{\kappa_1}}{\kappa_1-\kappa_2})e^{-\frac{\kappa_2}{2}(\tau_2-\tau_1)}\} \end{equation}
for $\tau_2\geq\tau_1\geq0$. Obviously the second photon cannot increase the decay rate if $\kappa_2<\kappa_1$ and so we only consider the case $\kappa_2>\kappa_1$. We aim to maximize the component of Eq.~(\ref{ste}) that decays with the rate $\kappa_2$, and minimize the component with the decay rate $\kappa_1$. This is done by letting
\begin{equation}\label{se4}
1+\frac{2\sqrt{\kappa_1}}{\kappa_1-\kappa_2}=0.
\end{equation}
In this case, the faster decay rate $\kappa_2$ dominates. The emission of the photon tends to synchronize with the faster decay rate of the second incoming photon, causing photon bunching in the output field. As a consequence, we will observe both photons in the output field earlier than the typical spontaneous emission time. The optimal $\kappa_2$ is thus given by
\begin{equation}\label{k2k1}
\kappa_2=\kappa_1+2\sqrt{\kappa_1}.
\end{equation}

Since $\xi^{'}(\tau_1,\tau_2)$ is a real function, we can conveniently plot $\xi^{'}(\tau_1,\tau_2)$ to compare the performance of different $\kappa_2$, see Figure~\ref{fig2}. We let $\kappa_1=1$. The maximal photon bunching is observed when $\kappa_2=3$. When $\kappa_2=10$, although there is higher probability that the stimulated emission would happen in a short time $(\tau_1,\tau_2<0.5)$, the probability of observing a delayed second photon is also high compared to $\kappa_2=3$ case, e.g., $(\tau_1\approx0.1,\tau_2=1)$. In other words, there is still significant probability for detecting the photon anti-bunching.

The previous works \cite{Rep12,Valente12} model the stimulated emission as a waveguide containing an incident photon in interaction with a two-level excited atom. The output states are then obtained using a real-space approach, that is, by solving the Schr\"{o}dinger equation for two-photon wavefunctions. In contrast, our approach does not rely on a stationary-state expansion and so the calculation is more straightforward. In particular, the optimal stimulated emission has been studied in \cite{Valente12} using the two-time correlation function of the output. As we have mentioned before, the correlation analysis can be easily done since we can obtain the exact output state. As a matter of fact, if the decay constant of the atom is normalized to $1$, then the optimal $\kappa_2$ would be $3$ according to the calculations in \cite{Valente12}, which is consistent with our result Eq.~(\ref{k2k1}).

In the above we have considered the optimal stimulated emission. The spontaneous emission with decay rate $\kappa_1$ will be enhanced if Eq.~(\ref{k2k1}) holds, and both photons will tend to decay with the faster rate $\kappa_2>\kappa_1$. Next we will discuss the suppression of spontaneous emission when $\kappa_2<\kappa_1$. Again considering Eq.~(\ref{ste}), we still need to maximize the component that decays with the slower rate $\kappa_2$. However, since $\kappa_1>\kappa_2>0$, Eq.~(\ref{k2k1}) is not feasible and so the optimal suppression of spontaneous emission cannot be realized. In order to prolong the life time of the excited atom, we need to consider alternative pulse function for the second photon. For example, we could define the piecewise pulse function $\xi_2(t_2)=\sqrt{\kappa_2/(e^{\kappa_2T}-1)}e^{\frac{\kappa_2}{2}t_2},t_2\in[0,T]$ and $\xi_2(t_2)=0$ elsewhere. Then we have
\begin{equation}\label{stenew}
\fl \xi^{'}(\tau_1,\tau_2)=\frac{\sqrt{\kappa_1\kappa_2}}{\sqrt{e^{\kappa_2T}-1}}e^{-\frac{\kappa_1-\kappa_2}{2}\tau_1}\{(1+\frac{2\sqrt{\kappa_1}}{\kappa_1-\kappa_2})e^{\frac{\kappa_2}{2}(\tau_2-\tau_1)}+(1-\frac{2\sqrt{\kappa_1}}{\kappa_1-\kappa_2})e^{-\frac{\kappa_1}{2}(\tau_2-\tau_1)}\} \end{equation}
for $T\geq\tau_2\geq\tau_1\geq0$. The condition to minimize the spontaneous emission is given by
\begin{equation}
1-\frac{2\sqrt{\kappa_1}}{\kappa_1-\kappa_2}=0,
\end{equation}
or
\begin{equation}\label{ope}
\kappa_2=\kappa_1-2\sqrt{\kappa_1},
\end{equation}
which requires $\kappa_1>4$. When we choose $\kappa_2$ to satisfy Eq.~(\ref{ope}), the first emission at $\tau_1$ is induced by the reduced decay rate $\kappa_1-\kappa_2$, and the spontaneous emission with the decay rate $\kappa_1$ is maximally suppressed before the emission of the second photon. Note that $\xi_2(t_2)$ is similar to the rising exponential pulse function which is designed to excite the atom.

\subsection{Two-channel and one-channel scattering}\label{twoone}
As we have mentioned, two coupling channels are used for the modelling of the left-going and right-going optical fields which are scattered at the emitter. The coupling operator is given by $L=[\sqrt{\kappa_1}\sigma_-\quad \sqrt{\kappa_2}\sigma_-]^T=[\sqrt{\kappa_1}\quad \sqrt{\kappa_2}]^T\sigma_-$. Let $L^{'}=\sigma_-$ and we can exactly calculate the output field state. Suppose the two input photons are separated in two channels, with $\xi_i(t),i=1,2$ being the pulse functions of the input photons travelling in the $i$-th channel. The output field state is expressed as
\begin{equation}
\fl |\Psi_{field}(\infty)\rangle=\int_{-\infty}^\infty\int_{-\infty}^\infty d\tau_1d\tau_2[\xi_{11}^{'}(\tau_1,\tau_2)(b_1^\dag)^2+\xi_{22}^{'}(\tau_1,\tau_2)(b_2^\dag)^2+\xi_{12}^{'}(\tau_1,\tau_2)b_1^\dag b_2^\dag]|0\rangle.
\end{equation}
There are three different two-photon components in the output field, namely, the probability of two photons in the first channel, two photons in the second channel, and one in the first and one in the second.

We consider the probability $P$ of observing at least one photon in the first output channel
\begin{equation}
P=\int_{-\infty}^\infty\int_{-\infty}^\infty d\tau_1d\tau_2(|\xi^{'}_{11}(\tau_1,\tau_2)|^2+|\xi^{'}_{12}(\tau_1,\tau_2)|^2).
\end{equation}
For simplicity we assume $\kappa_1=\kappa_2=\kappa$ which is the case for a waveguide system (the emitter couples to the optical fields with equal strength). The expressions for $\xi^{'}_{11}(\tau_1,\tau_2)$ and $\xi^{'}_{12}(\tau_1,\tau_2)$ with $\tau_1\geq\tau_2$ are calculated as
\begin{eqnarray}\label{xiex}
\fl\xi_{11}^{'}(\tau_1,\tau_2)=-\kappa\int_{-\infty}^{\tau_2}e^{-\kappa(\tau_2-s)}\xi_2(s)\xi_1(\tau_1)ds-\kappa\int_{-\infty}^{\tau_1}e^{-\kappa(\tau_1-s)}\xi_2(s)\xi_1(\tau_2)ds\nonumber\\
\fl+\kappa^2\int_{-\infty}^{\tau_1}e^{-\kappa(\tau_1-r)}dr\int_{-\infty}^{\tau_2}e^{-\kappa(\tau_2-s)}ds[\xi_2(s)\xi_1(r)+\xi_1(s)\xi_2(r)]\nonumber\\
\fl-2\kappa^2e^{-\kappa(\tau_1+\tau_2)}\int_{-\infty}^{\tau_2}e^{\kappa(2s+r)}ds\int_{-\infty}^sdr[\xi_1(s)\xi_2(r)+\xi_1(r)\xi_2(s)],\nonumber\\
\fl\xi_{12}^{'}(\tau_1,\tau_2)=-\kappa\int_{-\infty}^{\tau_2}e^{-\kappa(\tau_2-s)}\xi_2(s)\xi_1(\tau_1)ds-\kappa\int_{-\infty}^{\tau_1}e^{-\kappa(\tau_1-s)}\xi_1(s)\xi_2(\tau_2)ds\nonumber\\
\fl+\kappa^2\int_{-\infty}^{\tau_1}e^{-\kappa(\tau_1-r)}dr\int_{-\infty}^{\tau_2}e^{-\kappa(\tau_2-s)}ds[\xi_2(s)\xi_1(r)+\xi_1(s)\xi_2(r)]\nonumber\\
\fl-2\kappa^2e^{-\kappa(\tau_1+\tau_2)}\int_{-\infty}^{\tau_2}e^{\kappa(2s+r)}ds\int_{-\infty}^sdr[\xi_1(s)\xi_2(r)+\xi_1(r)\xi_2(s)]+\xi_1(\tau_1)\xi_2(\tau_2).
\end{eqnarray}
\begin{figure}
\includegraphics[scale=0.8]{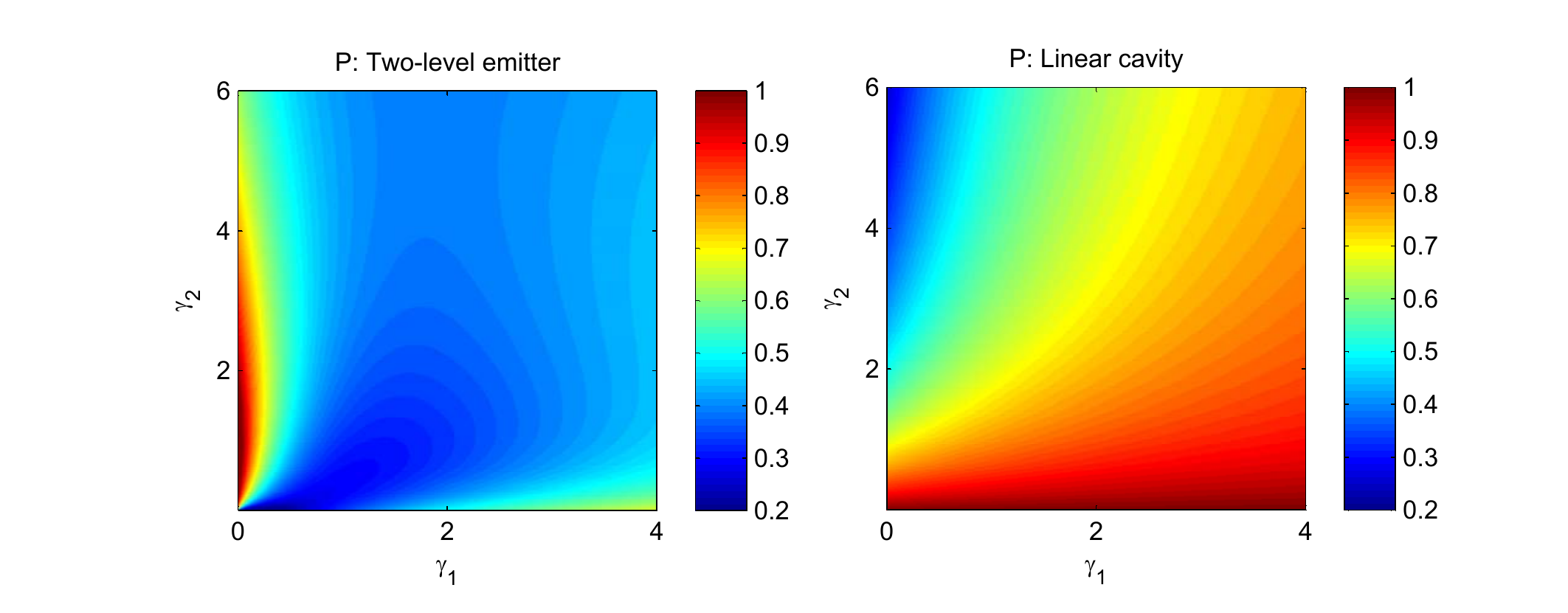}
\caption{Left: The transmission probability of a two-level emitter. Right: The transmission probability of a single-mode linear cavity in response to the same two-photon input Eq.~(\ref{fig2in}). The decay rate of the cavity mode is $\kappa$, which is the same as the decay rate of the emitter. There exists a rapid transition of $P$ for a two-level emitter when $\gamma_1<0.5$. }
\label{fig3}
\end{figure}
Suppose the pulse functions of the two input photons are given by
\begin{equation}\label{fig2in}
\xi_1(t_1)=-\sqrt{\gamma_1}e^{\frac{\gamma_1}{2}t_1}(1-u(t_1)),\quad \xi_2(t_2)=-\sqrt{\gamma_2}e^{\frac{\gamma_2}{2}t_2}(1-u(t_2)),
\end{equation}
with $\gamma_1,\gamma_2$ being controllable parameters. In this case, the relation between the transmission probability $P$ and the parameters $\gamma_1,\gamma_2$ is shown in Figure~\ref{fig3}. Also we have calculated the two-photon output state if the two-level emitter is replaced by a linear cavity. The first observation is that the responses of the linear and two-level system are quite distinct. The nonlinearity induced by the two-level emitter could significantly decrease the transmission probability in a two-photon process. This conversion of the transmission behaviour is consistent with the previous findings, e.g. in \cite{Shen07,Fan10,Liao13,Nysteen14,PhysRevLett.113.183601}. For instance, for a single photon incident on the emitter, the transmission probability goes to zero for long pulses and goes towards $1$ as the pulse width goes to $0$. For the latter case, the photon is no longer on resonance with the emitter. However, the presence of a second photon drastically changes the above behavior. This is because the two-photon state has different resonant frequencies compared to single-photon state. A two-photon state is in resonance if its energy matches the two-excitation eigenstate of the system \cite{Re12,Shi11}. As a two-level emitter can store at most one quanta, the photons cannot enter the emitter simultaneously. This leads to the nonlinear behaviour which is different from the single photon case. In addition, since photons do not interact in linear systems, the switching property of a linear cavity is the same for single-photon and two-photon states. Secondly, when we change the values of $\gamma_1$ and $\gamma_2$ for a linear system, the corresponding variation of $P$ is smooth. However for a two-level emitter, the variation of $P$ exhibits nonlinear behaviour when $\gamma_1$ is small, which is a signature of strong photon-photon correlation \cite{Shen07,Shi11,Liao13}. To be specific, if we fix a small $\gamma_1$, the transmission probability $P$ may fluctuate from the minimum value to the maximum value for a small variation of $\gamma_2$.

For the purpose of further illustrating the nonlinearity of two-photon interaction, next we consider the one-channel scattering by defining $L=\sqrt{\kappa}\sigma_-$. In the literature, the analytical study of the two-photon scattering often makes use of frequency-domain scattering analysis \cite{Fan10,Xu13,Xu15} or diagrammatic approaches \cite{PhysRevLett.113.183601}. Following the previous derivations, the time-domain pulse function $\xi^{'}(\tau_1,\tau_2)$ of the output state can be exactly calculated. Defining the Fourier transform of $f(t)$ as $f(\omega)=\int_{-\infty}^\infty f(t)e^{-2\pi\omega t\mbox{i}}dt$, the frequency-domain representation of $\xi^{'}(\tau_1,\tau_2)$ is given by
\begin{eqnarray}
\fl \xi^{'}(\omega_1,\omega_2)=\xi_1(\omega_1)\xi_2(\omega_2)+\xi_1(\omega_2)\xi_2(\omega_1)\nonumber\\
\fl-\kappa[\frac{1}{\frac{\kappa}{2}+2\pi\omega_1\mbox{i}}+\frac{1}{\frac{\kappa}{2}+2\pi\omega_2\mbox{i}}][\xi_2(\omega_1)\xi_1(\omega_2)+\xi_2(\omega_2)\xi_1(\omega_1)]\nonumber\\
\fl+\frac{\kappa^2}{(\frac{\kappa}{2}+2\pi\omega_2\mbox{i})(\frac{\kappa}{2}+2\pi\omega_2\mbox{i})}[\xi_2(\omega_1)\xi_1(\omega_2)+\xi_2(\omega_2)\xi_1(\omega_1)]\nonumber\\
\fl-\frac{4\kappa^2}{(\frac{\kappa}{2}+2\pi\omega_2\mbox{i})(\frac{\kappa}{2}+2\pi\omega_1\mbox{i})}\int_{-\infty}^\infty d\tau[\frac{\xi_2(\tau)}{\frac{\kappa}{2}+2\pi\tau\mbox{i}}\xi_1(\omega_1+\omega_2-\tau)+\frac{\xi_1(\tau)}{\frac{\kappa}{2}+2\pi\tau\mbox{i}}\xi_2(\omega_1+\omega_2-\tau)],\nonumber\\
\end{eqnarray}
where $\xi_1(\cdot),\xi_2(\cdot)$ are the pulse functions of the first and second photon, respectively. The last term in the above equation characterizes the inelastic scattering of two photons. To be more specific, the output photons with frequencies $\omega_1$ and $\omega_2$ can be generated by a pair of incident photons with different frequencies $\tau$ and $\omega_1+\omega_2-\tau$. This energy-preserving inelastic scattering property matches the previous findings in \cite{Fan10,Xu13,PhysRevLett.113.183601}.

\section{Conclusion}\label{seccon}
We have proposed a QSDE approach to model the two-photon scattering process for a general quantum system and calculate the time-domain response. We have studied only two cases in Section~\ref{AC} which enable the exact calculation of the two-photon output state. Nevertheless, there may exist quantum systems which allow exact analysis but do not belong to the two cases. For example, the response of a two-level system embedded in an open cavity or a linear cavity with Kerr nonlinearity has been analytically studied in \cite{Xu15,Re12}. The specific system operators considered in these works do not satisfy the sufficient conditions proposed in Section~\ref{AC}. However, it is easy to show that the coefficient terms of the output state can be exactly calculated based on a QSDE analysis. As a result, these systems are all amendable to the QSDE approach proposed in this paper. It will be interesting to investigate whether there exists a more general condition which implies the analytical computability of two-photon response for all these systems.

As pointed out in \cite{Rep12}, a photon must exist as a pulse in both waveguide and free-space. The analytical approach proposed in this paper is thus directly applicable to temporal pulse shaping \cite{Guofeng13}, as compared to the previous frequency-domain approaches. Moreover, the QSDE approach is extremely powerful in modelling a network of quantum systems, which has also been demonstrated in Section \ref{secpl}. Therefore, the analysis of two-photon response for a complicated quantum system could be benefited from this research. We expect that our results may be more general and directly applicable to the design of on-chip quantum circuit, in which the propagating photons could be scattered by linear and two-level components. For such practical systems, the QSDE approach and network theory may need certain extension in order to study arbitrary system parameters, relaxation and non-Markovian effects \cite{PhysRevLett.113.183601}.
\section*{References}

\bibliographystyle{unsrt}

\bibliography{two-photon}

\end{document}